\documentclass[journal=jacsat,manuscript=article]{achemso}

\usepackage[version=3]{mhchem} 
\usepackage{chemformula} 
\usepackage[T1]{fontenc} 

\author{Ashly Sunny}
\affiliation{Department of Physics, School of Advanced Sciences, Vellore Institute of Technology, Vellore, Tamilnadu - 632014, India}
\author{Aniket Balapure}
\affiliation{Department of Chemistry, Birla Institute of Technology and Science (BITS), Pilani, Hyderabad Campus, Jawahar Nagar, Kapra Mandal, Medchal District, Hyderabad, Telangana – 500078. India}
\author{Ramakrishnan Ganesan}
\affiliation{Department of Chemistry, Birla Institute of Technology and Science (BITS), Pilani, Hyderabad Campus, Jawahar Nagar, Kapra Mandal, Medchal District, Hyderabad, Telangana – 500078. India}
\author{R. Thamankar}
\affiliation{Centre for Functional Materials, Vellore Institute of Technology, Vellore, Tamilnadu - 632014. India}
\email{rameshm.thamankar@vit.ac.in}
\phone{+91 9742430830}
\email{rameshm.thamankar@vit.ac.in}

\title[]
  {Room temperature deep UV photoluminescence from low dimensional hexagonal boron nitride prepared using a facile synthesis }

\abbreviations{PL, XRD, UV }
\keywords{Photoluminescence, defects, hBN,}

\begin{document}




\begin{abstract}
  Evaluation of the defect levels in low-dimensional materials is an important aspect of quantum science. In this article, we report a facile synthesis method of hexagonal boron nitride (h-BN) and evaluate the defects and their light emission characteristics. The thermal annealing procedure is optimized to obtain clean h-BN. The UV-Vis spectroscopy shows the optical energy gap of 5.28 eV which is comparable to the reported energy gap for exfoliated, clean h-BN samples. The optimized synthesis route of h-BN has generated two kinds of defects which are characterised using room temperature photoluminescence measurements. The defects emit light at 4.18 eV (in deep ultraviolet region) and 3.44 eV (ultraviolet), respectively. The defect emitting deep ultraviolet (DUV) has oscillatory dependency on the excitation energy, while that emitting 3.44 eV light (ZPL3.44 eV) has a phonon bands with mean energy level separation of 125 meV measured at room temperature. This agrees very well with the Franck-Condon-like structure having regularly spaced energy levels, which are typical indications of single defect levels in the low dimensional h-BN.

\end{abstract}

\section{Introduction}
Ever since graphene is exfoliated from graphite, curiosity in low dimensional materials has been exponential and efforts in understanding the physics in low dimensions has regained importance\cite{Neto2009, Novoselov2004,Geim2007}. Various low dimensional materials have been studied which are stable at room temperature\cite{Mak2010, Bhimanapati2015}. These materials show thickness dependent physical properties such as electronic band structure and energy gap as in the case of MoS$_{2}$ and show quantum size effects. Graphene as such lacks electronic energy gap, which is required for suitability in the electronic industry. MoS$_{2}$ is another two-dimensional material studied extensively in the last two decades, which has an intrinsic electronic energy gap down to monolayer. Single layer of MoS$_{2}$ shows direct energy gap of 1.8 eV and indirect energy gap of 2 eV \cite{Mak2010,Splendiani2010,Eda2011}. The electronic properties are completely different for a monolayer and multilayered materials. Since materials like MoS$_{2}$ can be exfoliated easily, much of the effort has been to use them in fabrication of nanoelectronic devices. Tremendous research has been done to use such monolayers to fabricate hetero structures with various other low dimensional materials and thin films\cite{Novoselov2016,Gibertini2019}. From a no-gap graphene to moderate energy gap materials such as MoS$_{2}$ and chalcogenides have been studied for a decade now\cite{Radisavljevic2011,Gomez2015,Wang2012,Sangwan2018}. It is a well-established fact that majority of these low dimensional materials show a change in the electronic band structure with number of layers resulting in direct-indirect band gap transition. Materials like monolayer MoS$_{2}$ and InSe show direct band gap around 1.8 eV, while the energy gap reduces to bulk value of about 1.2 - 1.25 eV when measured for 2-4 monolayers \cite{Splendiani2010,Sun2018}.

\begin{figure}[t]
\centering
\includegraphics[width=12cm]{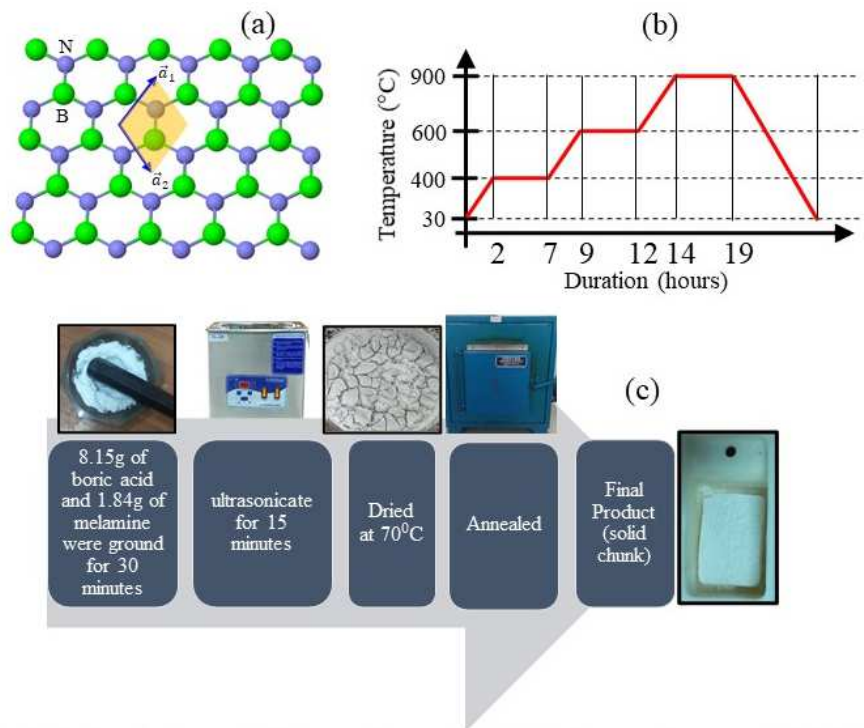}
\caption{(a) A schematic diagram of 2D hexagonal boron nitride (h-BN). Honeycomb structure of 2D h-BN with Bravais lattice vectors a$_{1}$ and a$_{2}$. The ball-stick model represented by boron atoms(green) and nitrogen atoms (blue). (b) The optimized heat treatment involved in the synthesis of clean h-BN. Both the axes are not to the scale. ( c)The synthetic route of h-BN. The samples are annealed in a box furnace in ambient conditions. The h-BN has appearance of bright white solid chunk as shown.}
\end{figure}
Recently, the two-dimensional wide band gap materials have also attracted lot of attention. Materials such as hexagonal boron nitride (h-BN), Silicon Carbide (SiC) and diamond(C) has attracted much attention due to their extremely large electronic energy gap at monolayer limit ($\sim$ 6eV) and also stability at room temperature\cite{Kubota2007,Museur2008,Watanabe2011,Baranov2011,Koehl2011,Balasubramanian2009, Dzurak2011}. In particular, the hexagonal boron nitride (h-BN) shows layered structure similar to graphene. In view of this, h-BN can be considered as a versatile material for electronic devices. h-BN shows hexagonal lattice with an alternative boron-nitrogen arrangement within the lattice, which is quite similar to the atomic arrangement of graphene. It is quite fascinating that even though h-BN and graphene(C) show similar atomic arrangements (the nearest neighbour distance, lattice parameters and interlayer spacing), h-BN shows a large energy gap ($\sim$6 eV), while 'pure' graphene has no energy gap\cite{Pakdel2012}.
Majority of the studies performed on h-BN is by using bulk single crystal and mechanically exfoliating single to few layers of h-BN. Despite the extensive research performed on this material, the fundamental question about the energy gap remains still debatable\cite{Cassabois2016,Watanabe2004}. The \emph{ab-initio} calculations and photoluminescence experiments give quite contrasting interpretations with large scatter in the values for the energy gap (E$_{g}$). While the \textit{ab-initio} calculations predicting an indirect band gap, the optical measurements indicate a direct band gap\cite{Xu1991,Furthmueller1994,Blase1995,Arnaud2006,Evans2008}. The electronic band structure undergoes a crossover from indirect band gap (bulk h-BN) to direct band gap (monolayer h-BN)\cite{Elias2019}. In this context, h-BN is a wide band gap layered material which hosts extremely bright light emitting defect centres stable at room temperature \cite{Tran1,Tran2}. The defects present in these two dimensional materials emitting single photon has been studied recently. This is a primary requirement for single photon emitting source and quantum photonics \cite{Tran1,Toth2019,Gil2020,Exarhos2019}.
Theoretical investigations on monolayer hexagonal boron nitride reveals overestimated wide energy gap compared to experimentally observed value. The energy gap calculated using the density functional theory (DFT) calculations by the Bethe-Salpeter equation is more than 7 eV\cite{Ferreira2019,Ba2017}. On the other hand, there are reports about calculations giving 6.47 eV using DFT-VASP, yet there is no agreement between various procedures adopted in the calculations. Recently, experiments performed on h-BN/Graphite using scanning tunneling microscopy/spectroscopy (STM/STS) suggested an energy gap of 6.8 $\pm$ 0.2 eV \cite{Roman2021}. The optical energy gap calculations of monolayer h-BN are determined by the exciton energy and a range of energy gap values from 5.30 – 6.30eV have been calculated \cite{Ferreira2019}.
h-BN is also successfully synthesized using a variety of other techniques such as atomic layer deposition\cite{Park2017}, chemical vapour deposition\cite{Kim2012} ,molecular beam epitaxy \cite{Pierucci2018,Elias2019} and layer-by-layer sputtering process\cite{Jin2009}.
The low dimensional wide band gap materials not only find their importance in the nanoscale transistor design, but also for hosting the fundamental unit of quantum information processing i.e. qubit\cite{Weber2010}. In this case, a two level quantum system is idealised and a transition between these two energy levels will emit a photon of a given wavelength. An ideal two level system is provided by an isolated defect in 2-dimensional material having large energy gap. Two-dimensional (2-D) systems which are structurally open, demonstrating the quantum confinement, and then reduced charge screening are ideal qualities for hosting the single photon emitters (SPE's). Since the electronic states of the single defect is well separated by the energy bands of the host lattice, it can be controlled by using the polarized light\cite{Gupta2019,Frey2020}.
For example, the colour centres in nano-diamonds are one such material platform which are stable and give good emission\cite{Wolters2010}. Similarly, the h-BN nanoparticles can host desirable defect centres which could be used in devices for quantum technologies.
Interestingly, even a medium energy gap 2-dimensional material such as MoS$_{2}$ is also capable of hosting isolated defects. In a recently published article, it has been proved that sulphur defects can be prepared controllably using He-ion radiation which act as single photon emitter\cite{Klein2021}. In essence, these defects are excellent candidates for quantum emission. The advancement in the techniques used in miniaturisation of electronic devices has led to conceive these atomic scale devices using either atoms or molecules.
\begin{figure}[h]
\centering
\includegraphics[width=12cm]{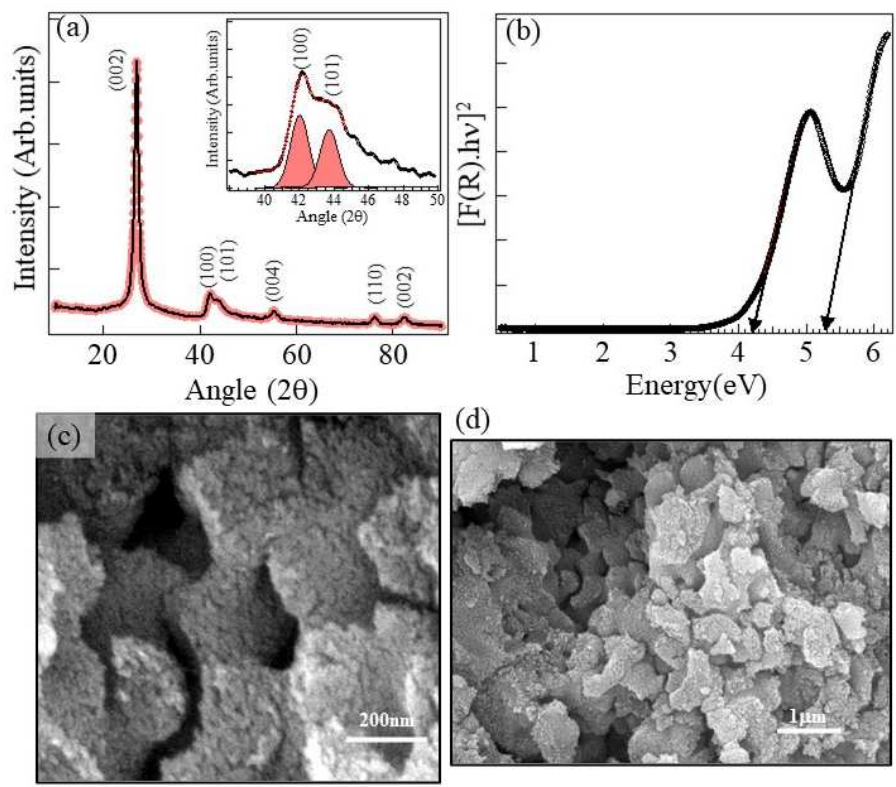}
\caption{(a)A typical X-ray diffraction pattern of multilayer hexagonal boron nitride. The principal reflection is at 26.8$^\circ$ typical of the (002) reflection in h-BN. The inset in (a) shows peaks resolved appearing at 42$^\circ$/43.7$^\circ$ showing (100)/101 reflections. (b) the Tauc plot to calculate the optical band gap. The two band edges corresponding to 5.28eV and 4.2eV are shown. (c) and (d) are high resolution scanning electron micrographs showing large area h-BN formed.
}
\end{figure}
The possible use of h-BN in the graphene based transistor design is very promising. Firstly, there is a good lattice match resulting in a low strained heterostructures and h-BN showing large band gap is essential as a dielectric in the transistor design. There have been numerous reports on the mechanism of dielectric breakdown of a few layer h-BN using local breakdown studies. Notably, the nanoscale measurements using atomic force microscopy/spectroscopy are very crucial for understanding the fundamental mechanism of dielectric breakdown\cite{Ranjan2021}.

We adopted a simple facile procedure to synthesize multilayered h-BN. Herein, we report the fundamental analysis of prepared h-BN multilayers using X-ray diffraction, UV-Vis spectroscopy, and X-ray photoelectron spectroscopy (XPS). Further, detailed photoluminescence studies been performed at room temperature to understand the defect energy levels and their light emitting properties.

\section{Results and discussion}

\subsection{Experimental Details}

\begin{figure}[!]
\centering
\includegraphics[width=12cm]{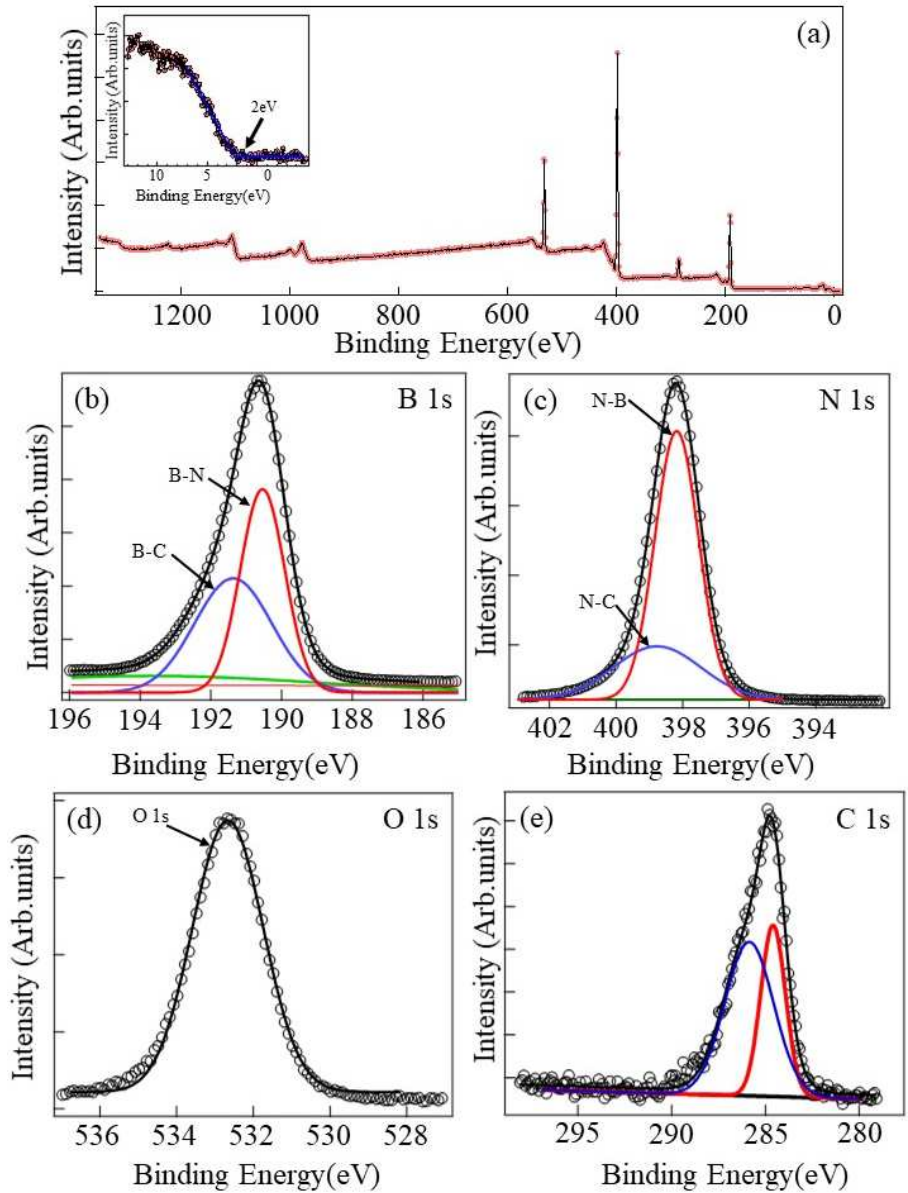}
\caption{XPS measurements on h-BN samples. (a) Survey scan of h-BN sample. The peaks for oxygen and carbon are denoted. The valance band maximum is shown in the inset. The line scans meet at 2eV below the Fermi energy. (b)  The high resolution scan of boron. Peak is at 190.56 eV correspond to B-N and the small shoulder at 191.56 eV correspond to B-C. (c) high resolution N-peak. The main peak is at 398.15 eV which can be attributed to N-B and the shoulder at 399.2 eV can be attributed to N-C. (d)  high resolution peak at oxygen binding energy. (e) high resolution C1s peak at 284.59 eV and a shoulder at 285.56 eV.}
\end{figure}

In a typical synthesis procedure as shown in Fig (1), 8.15g of boric acid and 1.84g of melamine (in total 10g of the initial precursors by weight) are taken and mixed well using mortar and pestle for 30 min and then transferred to a beaker containing methanol. This precursor mixture was ultrasonicated for two hours and then dried at 70$^\circ$C, refer Fig. (1c).  Once the material is completely dried, it is transferred to a Alumina boat and placed in a furnace in an ambient condition. We have tested various heating cycles considering the evolution of nitrogen from the melamine source, desorption of oxygen from the materials at different temperature. The heating cycle which resulted in pure hexagonal boron nitride is shown in fig.(1b). The sample is heated with a heating rate of 6$^\circ$C/min to 400$^\circ$C and kept it for 5 hours. Here, the melamine molecules partially dissociate into a nitrogen rich precursor for the h-BN. In the second stage, the temperature is raised to 600$^\circ$C and kept for 3 hours.  In this step, the boric acid partially dissociate to give rise to a boron rich precursor for h-BN. In the third stage, the temperature is taken to 900$^\circ$C and kept for 5 hours. At this temperature, boron- and nitrogen-rich precursors will react to give the boron nitride. Here the material is kept for sufficient duration to complete the reaction and then allowed to cool down naturally to room temperature. The appearance of the h-BN immediately taken out of the furnace is shown in the fig.(1e) appears as a white solid chunk. The actual reaction mechanism is shown in fig.S1(sup. information)

The phase purity and crystalline structure of the synthesized material is examined using X-ray diffraction(XRD). We use CuK$\alpha$$_{1}$ radiation with wavelength of 1.504{\AA} and the data was collected from 10 - 90$^\circ$(2$\theta$) with a scanning rage of 2$^\circ$/min. Fig.(2) shows the XRD pattern obtained for the sample prepared with optimized annealing conditions. The principle diffraction peak is located at 2$\theta$ $\sim$ 26.9$^\circ$ which is assigned for the (002) peaks of the h-BN\cite{Li2016,Mato2016}. This peak pattern indicates that there is a similarity between graphite like h-BN structure.

Further, much weaker peaks are observed at 42$^\circ$ and 43.7$^\circ$ from the (100)/(101) in close agreement with the reported values in the literature. The broad peaks at this position represents stacking faults in the (100) planes\cite{Paine1990}. Such broad peaks are attributed to the ‘turbostatic’ structure mentioned in the literature. However, the XRD peaks at this diffraction angle can be clearly resolved which is close to the hexagonal boron nitride as shown in the inset Fig.(2a). Further, the peaks [55.25$^\circ$, (004)], [76.3$^\circ$, (110)] and [76.3$^\circ$, (002)] are indicative of a clearly formed single crystalline h-BN. The interlayer spacing of the (002) crystallographic planes is calculated using Bragg’s law and we obtain a value of 0.331 nm which is very close to the ideal value of 0.333 nm for hexagonal boron nitride. Corroborating our XRD analysis, the high resolution electron microscopy images confirm large area layers of h-BN in the form of multilayers. As shown in Fig.(2c and d), large area h-BN layers are formed. At this point, we are not able to separate them as single sheets. Efforts are underway to separate these sheets resulting into single sheets.

Next step is to determine the optical band gap and we used the standard UV-vis spectrometer to determine the absorbance, reflectance of the light by dried h-BN sample. The sample was ultrasonicated for prolonged duration before drying. We use the absorbance to calculate the optical energy gap by means of Tauc plot. As can be seen in fig.(2b), the calculated energy gap is about 5.28 eV which agrees very well with the reported energy gap values\cite{Lee2017}. It is to be noted that there is a large scatter in the determination of the optical band gap of h-BN in the literature. The energy gap measured in our samples match very well with the values obtained using UV-vis spectrum available in the literature\cite{Lee2017}. Additionally, we also can see a strong band of energy lying about 4.2 eV from the valence band edge.  This is attributed to the defects which are discussed in detail in the later sections.

To find out the chemical nature of the individual elements in h-BN multilayers, survey scan and the narrow energy XPS scans are performed around the binding energies of boron (B), nitrogen(N), carbon(C) and oxygen (O). Global spectra of h-BN is shown in the Fig.(3a) demonstrates a typical spectrum with narrow peaks at various binding energy values characteristic of high quality sample. The survey scan depicts binding energy peaks at 190.56 eV, 284.59 eV, 398.15 eV and 532.64 eV correspond to B(1s),C(1s),N(1s) and O(1s) from the h-BN sample respectively.
Firstly, the valence band maximum (VBM) can be calculated by performing a high-resolution scan near the Fermi energy. As shown in the inset of Fig.(3a), the VBM was calculated by considering the intersection of the flat XPS energy approaching the valence band(VB) of h-BN and the linear fit of the valence band edge in the XPS data. Such a calculation give us the VBM at 2.0 eV below the Fermi energy.  This is in very good agreement with the angle resolved photoemission spectroscopy (ARPES) measurement on h-BN/(graphite), h-BN/Ir(111)and h-BN single crystal where the VBM has been found at 2.2 - 2.8 eV below the Fermi energy. This implies that our facile synthesis method has resulted in a very good quality hexagonal boron nitride.
We take the binding energy of carbon (C 1s) at 284.59 eV as shown in the inset of Fig.(3b) as reference in all our analysis. The main peak is at 284.59 eV arising from the sp$^2$ hybridization of carbon coming from the carbon support. The small shoulder peak arising at 285.56 eV can be attributed to C-O bond which could be from the adsorbed oxygen on the surface of h-BN\cite{Balapure2020}. The most dominant peak for boron is $\sim$ 190.56 eV which is very close to the ideal value of boron position in hexagonal boron nitride \cite{Moulder1992}. This peak corresponds to the B-N bond in h-BN.  The maximum intensity of B-N peak at 190.56eV shows the formation of hexagonal phase of boron nitride\cite{Moulder1992,Marini2006,Marom2010}. A small shoulder can be seen in higher energy at 191.56eV representing the B-C bond.
Even though the survey scan does show a strong oxygen peak, it is not bonded to the boron atom giving a B-O binding energy at 192 eV \cite{Sainsbury2012}. From this it is clear that the oxygen peak shown in the full scan arises from the residual gas which is adsorbed on the surface\cite{Bhimanapati2014}. As reported in the results from various groups, the enhanced peak at 532.8eV attributed to the boron – oxygen bonding (corresponding to B$_{2}$O$_{3}$) is not seen in our sample, thereby reconfirming a complete hexagonal boron nitride phase formation in our sample.
The high resolution binding energy scan of nitrogen (N 1s)occurs at 398.15 eV correspond to the N-B bonding and a small shoulder exists at  399.2eV which can be attributed to N-C bond. Interestingly, the binding energy scan of oxygen shows up at 532.67 eV which is due to the presence of B-O bond. We assign this peak to the bond formation during the ultrasonication of the h-BN layers. This can indicate the presence of numerous functional groups on the h-BN surface.

\begin{figure}[!]
\centering
\includegraphics[width=12cm]{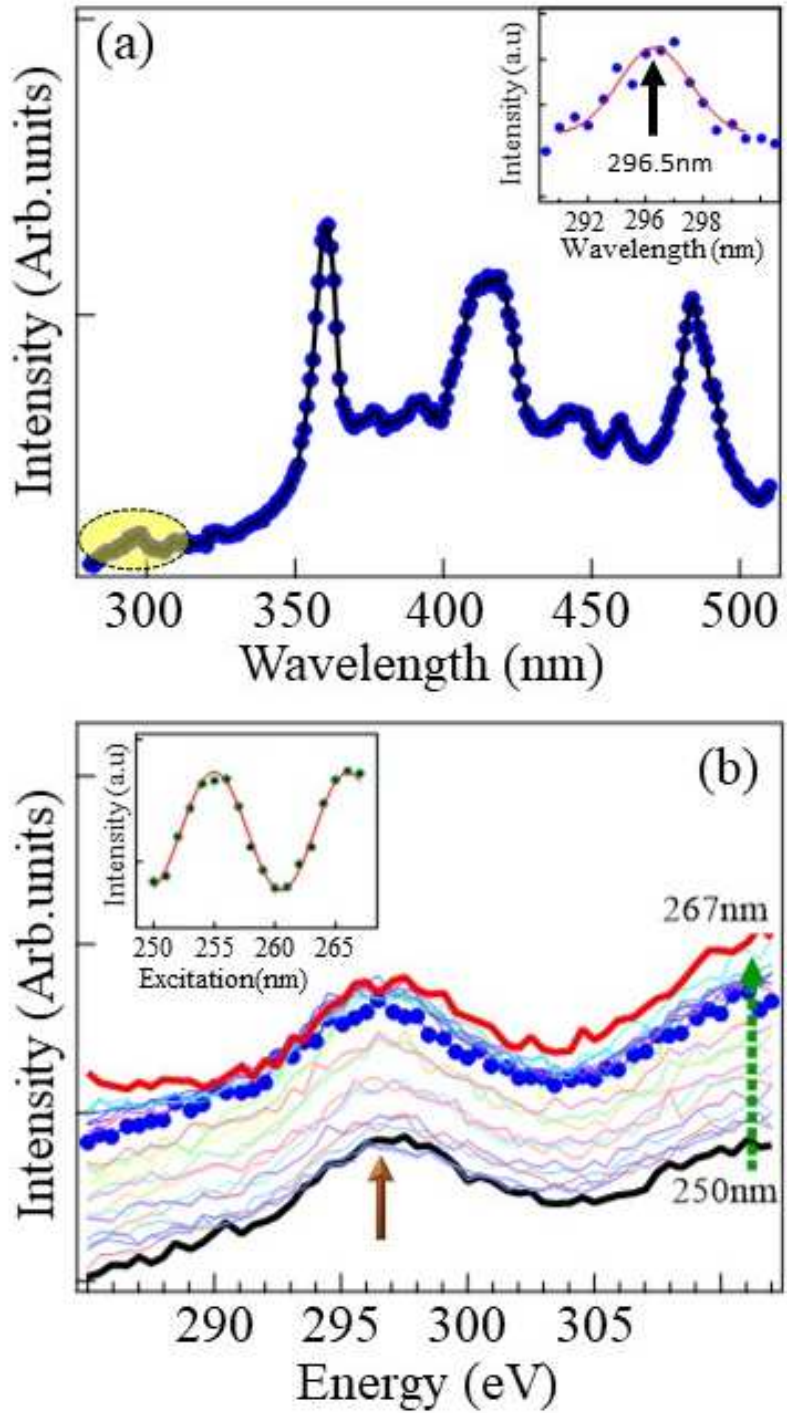}
\caption{(a) A typical photoluminescence spectra from h-BN dispersed in water. The excitation wavelength is 265nm.  A series of defect induced emission peaks are observed. The smallest wavelength (deep-UV) emission is observed at 296.5 nm [shaded region in the (a)]. The Inset in (a) shows a clear emission peak centred at 296.5 nm (4.18eV). (b) The excitation wavelength dependence of the 296.5 nm(4.18 eV) peak. The Black and red curves are the spectra obtained with 250 nm and 267 nm respectively. The excitation wavelength is varied and 296.5 nm(4.18 eV) peak shows a particular intensity variation as shown in the inset (b)}
\end{figure}

Careful analysis of the XPS spectrum and the BE position of various peaks reveal rich information about the defects present in h-BN. To support this, photoluminescence measurements has been a usual method employed to identify atomic scale defects in h-BN\cite{Museur2008}. Recently, using the Density functional theory (DFT), \textit{ab-initio} calculations and core level binding energy calculations, the deconvoluted broad XPS spectrum positions were used to assign certain kind of defects in the h-BN lattice \cite{Khorasani2021,Attaccalite2011}. Further, there are variety of defects observed in h-BN such as a neutral nitrogen vacancy (V$_{N}$), single electron trapped nitrogen vacancy (V$_{N}$)$^{-1}$, a carbon atom replacing nitrogen (C$_{N}$) and some complex defects such as N$_{B}$V$_{N}$ \cite{Sajid1,Sajid2,Hamdi2020}. Most of the time, a nitrogen vacancy is either neutral or single occupied (V$_{N}$) are predominant in h-BN samples which can be categorised as process induced defect is seen.

In order to check if our synthesis procedure resulted in such defects and decipher their light emitting properties, room temperature photoluminescence (PL) spectra was taken. As usual, the h-BN multilayered sample was excited using deep-UV light and check the emission lines. Since our sample contains multilayered h-BN, we expected luminescence from various defect levels present in the sample. In particular, when excited with 265nm(4.7 eV) photons and we achieved well resolved photoluminescence spectral. The selection of excitation wavelength(energy) is based on our Tauc plot where we have seen an additional band edge at 4.2 eV (see Fig.2b). A typical photoluminescence spectra is shown in Fig.(4a). A series of emission lines varying in energy position can be seen. Since our first interest is to understand the additional band edge observed in Tauc plot, the scan range in PL is restricted to small wavelength range up to 510 nm. Firstly we will discuss the PL spectra at lower wavelength (DUV) as shown in Fig(4a) (highlighted in yellow). For clarity, this is plotted in the inset of Fig.(4a). The Gaussian fit of the data in this wavelength range shows the peak position at 296.5 nm (4.18 eV). This PL intensity is often attributed to the presence of carbon substituted Nitrogen (V$_{N}$) vacancies\cite{Weston2018}. There are also arguments about the origin of this peak due to intrinsic origin rather than an extrinsic \cite{Tsushima2018}. The density functional theory (DFT) within the LSDA and GGA show that these levels are extended energy states close to conduction band edge\cite{Attaccalite2011}. The calculations also suggest that the observed peak can also be from the boron vacancy (V$_{B}$). Even though we are not able to resolve this energy state at this moment, but it strongly indicates that our sample preparation route has yielded very good h-BN samples. The low intensity of this peak suggests that the overall PL intensity from our sample is not from the V$_{N}$, but from other defects. A clearly resolved spectra in these DUV wavelength range is shown in Fig.S2(in supp. information).

We measured the effect of excitation wavelength (energy) on this emission line. The wavelength was varied from 250 nm(4.96 eV) to 272 nm(4.55 eV)as shown in Fig.4b. The black(red) curves represent the emission intensity with excitation 250 nm(4.96 eV) to 272 nm(4.55 eV) respectively and the blue dotted spectra is for the excitation wavelength 265nm. The measured intensity variation of PL at 296.5nm with excitation wavelength is plotted is shown in the inset Fig.(4b). The intensity has an oscillatory dependence on excitation wavelength. Further studies are needed to elucidate the excitation energy dependence of intensity variation in h-BN. This variation can be seen as an evidence for the bell shaped energy levels/bands participating in the excitation and emission from the defects \cite{Gribkovskii1998}. In order to check if the emission lines observed are due to solvent used, we performed the experiment using methyl alcohol as a solvent and compared the emission with that when water is used as a solvent. As shown in Fig. S3( in supp. information) both emission specta match very well thus confirming that all the spectral lines discussed here are from the hexagonal boron nitride.

To further analyse this spectrum in the longer wavelength regime (up to 510 nm), we plot the PL intensity variation with the incident photon energy as shown in the Fig.(5a). Apart from the deep ultraviolet (DUV) emitting  colour centre, there are series of emission lines in the energy range 2.4 to 3.45eV. We consider the sharp spectral line at 3.44 eV as a zero phonon line (ZPL) and what follows, we refer this to ZPL3.44 band and all the spectral features lie in the near-UV (NUV) region. similar near-UV bands for h-BN are reported\cite{Tsushima2018}. As can be seen in Fig.5a, the series of spectral lines which can be treated as the phonon bands associated to the ZPL3.44. The position of the peaks in this phonon band shows that they are equally spaced in the energy. We use the Gaussain peak fitting for each peak as shown by the coloured gaussian noted with numbers from 0 to 7. Overall intensity variation can be satisfactorily explained with this (black spectral curve). One should note that all the spectral peaks within the phonon bands do not have the same intensity and FWHM value. Further, to understand the spectral features, the peak position with respect to the ZPL3.44 peak are shown in the top x-axis. It is clear that the spectral features (phonon bands) are equally spaced with respect to the ZPL3.44.
Apart from the ZPL3.44, we see additional spectral features with equal separation within the experimental resolution. We plot the peak position E$_{max}$ value thus obtained as a function of peak number (n = 0,1,2,3...) as shown in Fig.(5b). The dependencies of E$_{max}$(n) with the peak number falls on a straight line indicating that these spectral features are indeed the phonon modes and the slope indicates the phonon energy h$\omega_{LO}$=125meV for the contributing defect in our sample\cite{Reich2005,Berzina2016}. These obtained phonon energies match very well with reported values confirming the emission from the single defect.

\begin{figure}[!]
\centering
\includegraphics[width=18cm]{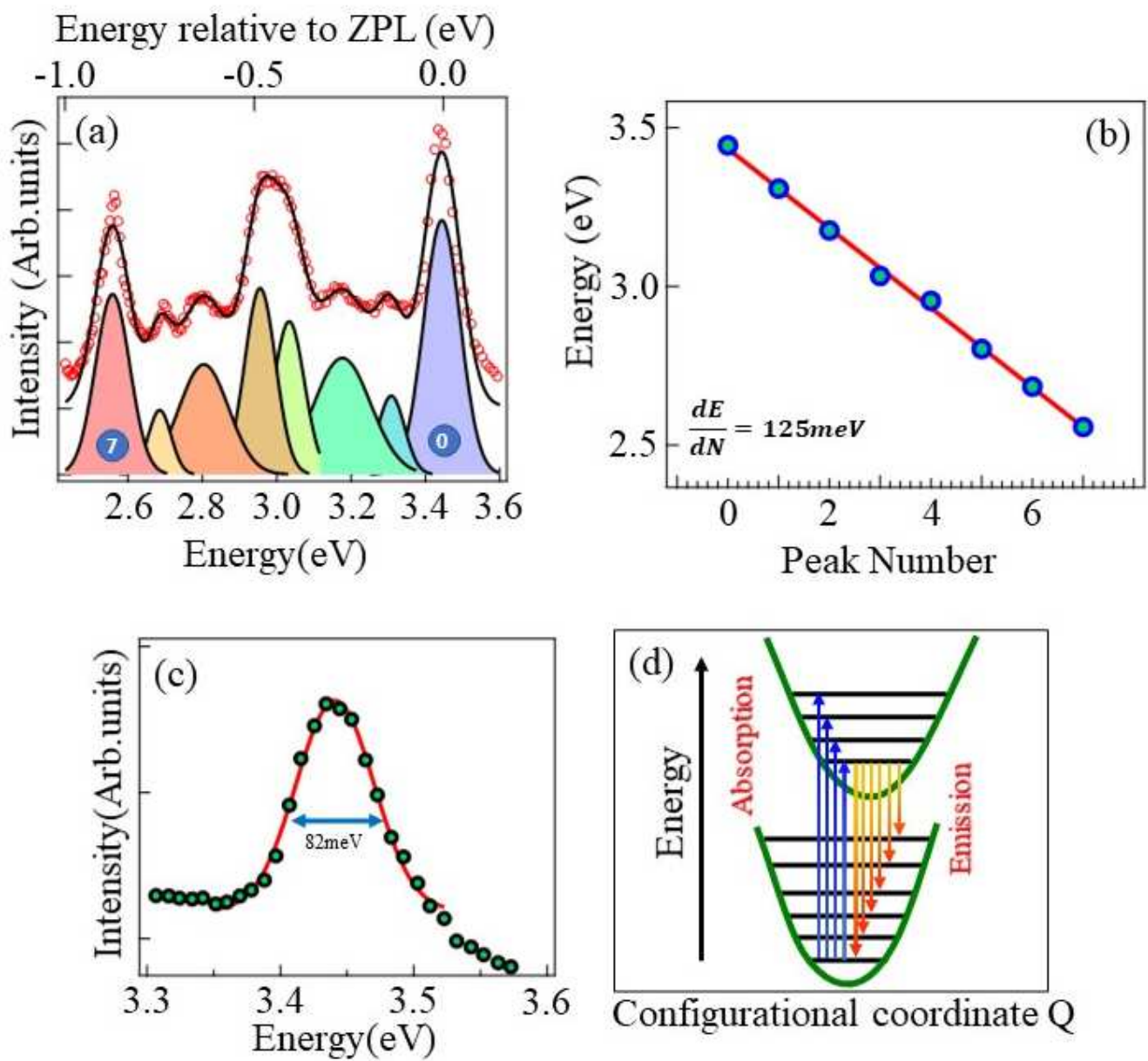}
\caption{(a) Re-plotting the photoluminescence spectra from h-BN dispersed in water (as shown in Fig.4a. The spectra is plotted again the energy along X-axis. A series of defect induced emission peaks are observed. The emission at 3.44 eV is taken as zero phonon line. Correspondingly other series of peaks are plotted with reference to ZPL (top axis of (a). These spectral peaks are satisfactorily fitted with gaussian fitting from (0) to (7). (b). (b) Energy positions of the spectral peaks are plotted show that they are equally spaced. The average spacing between the spectra is $\sim$ 125 meV. (c) The ZPL at 3.44 eV is plotted showing very good FWHM of $\sim$ 82 meV measured at room temperature. (d) The location of the spectral lines can be modelled using Franck-Condon type transition.}
\end{figure}

The optical response and the characteristic spectral lines shown from such defects can be modelled using the Franck - Condon (FC-type) type of spectrum. It is well known that in the case of an atomic systems or molecular system, the localised energy levels can strongly couple with optical phonons thereby causing the excitation (absorption) and de-excitation (emission) processes. The excitation caused due to incident light transfers electrons to the excited state while simultaneously annihilating and creating phonons. Such an interpretation has successfully explained the optical transitions between the defect states, notably the nitrogen vacancy (NV center) in diamond. The electronic excitation can couple with the phonons giving rise to phonon side band. The occurrence of phonon side band can also due to the localised modes of the defects in the lattice. Our measurements are done at room temperature, so the spectral features observed will be broad. In the FC-type molecular excitation, the highest energy transition which does not involve phonons is considered as zero-phonon line (ZPL) and in our case it is at 3.44 eV. The asymmetric zero phonon line is plotted in Fig.(5c) shows a very sharp intensity even though measured at room temperature. The FWHM matches very well with the measured values at low temperatures.
One of the methods in which the FC-type defect emission is characterised is by calculating the Debye - Waller(DW) factor and Huang-Rys (HR) factors. The DW factor is experimentally found by using the photoluminescence spectrum. It is calculated by taking the ratio of the ZPL and the overall spectrum. Debye - Waller factor is a measure of the electron - phonon coupling strength in the material. From the ZPL3.44 peak, the Debye - Waller factor can be calculated using spectral weight of the ZPL peak and overall emission in the ZPL3.44 band. We obtain a DW factor as 22\%. this is reasonable as our measurements are done at room temperature and the calculations of DW factor will contain the thermal broadening of the ZPL3.44 peak which will tend to result in a larger ratio. These quality factors are fairly good considering the procedure adopted for synthesis of the h-BN layers.  The corresponding Huang-Rys factor S = -$\ln$(w) is about 2.08$\pm$ 0.1. The calculation of the RH factor based on our room temperature measurements agree very well with the low temperature measurements.

\subsection{\textsc{conclusions}}
Understanding the defect structure in h-BN paves the way to fabricate single photon source with well defined emission wavelengths. The synthesis of such a photon source should be easy and cost effective. Here, Multilayer hexagonal boron nitride (h-BN) is prepared using a facile synthesis method which involves heating the precursors in an ambient conditions. A detailed analysis of the XRD pattern reveals that the optimized heating procedure indeed yielded a clean h-BN. Further, the UV-vis spectra shows energy gap (~ 5.28 eV) comparable to the reported values in literature. The process induced defects show light emission in the DUV region which is also very essential for the quantum communication in DUV region. We find two type of defects in our samples emitting a DUV light at 4.18 eV and UV light at 3.44 eV. We attribute these two defects to the nitrogen vacancies created during the synthesis of h-BN. The defect with ZPL at 3.44 eV shows a very sharp spectral line even at room temperature. This is associated with a phonon band with equally separated in energy levels. These spectral lines can be explained by considering the excitation-deexcitation process using the Franck - Condon principle. Harmonic oscillator approximation is utilised and the energy level separation is found to be about 125 meV at room temperature. The quality factors are calculated by using the cumulative areas of the spectral lines and the quality factors match very well with that of exfoliated h-BN layers. Our synthesis method is an easy-to-do route to prepare high quality h-BN with deep UV emitting defects, which are useful in the DUV-photonic applications and quantum computation.

\begin{acknowledgement}

Ashley Sunny and R.Thamankar would like to thank Vellore Institute of Technology for their support during this research work.
R.Thamankar acknowledges the support of Dr. RG from BITS Hyderabad for the XPS measurements.
RT would like to acknowledge Dr. Rajagopal Department of Chemistry, VIT Vellore for critical reading of the manuscript.

\end{acknowledgement}

 Information, which should use the
\begin{suppinfo}

\begin{itemize}
    \item Figure S1.: \emph{The reaction scheme of hexagonal Boron Nitride preparation}.
    \item Figure S2.: \emph{The excitation wavelength dependence of PL at lower wavelength region}.
    \item Figure S3.: \emph{Emission spectra using two solvents, H$_{2}$O and CH$_{3}$OH. }.
\end{itemize}

\end{suppinfo}

\bibliography{Manuscript.bbl}

\end{document}